\documentclass[reprint,aps,superscriptaddress,floatfix,amsmath,amssymb,showpacs,nofootinbib]{revtex4-1}
\usepackage{hyperref}
\usepackage{microtype}

\newcommand{\ket}[1]{\ensuremath{|{#1\rangle}}}

\newcommand{\ketbra}[2]{\ensuremath{|{#1 \rangle}{\langle #2}|}}
\newcommand{\op}[1]{\hat{#1}}

\begin{document}

\title{What classicality? Decoherence and Bohr's classical concepts}

\author{Maximilian Schlosshauer}

\email{schlosshauer@nbi.dk}

\affiliation{Niels Bohr Institute, University of Copenhagen, Blegdamsvej 17, 2100
    Copenhagen, Denmark}

\author{Kristian Camilleri}

\affiliation{School of Philosophy, Anthropology, and Social
    Inquiry, University of Melbourne, Melbourne, Victoria 3010, Australia}



\begin{abstract}
Niels Bohr famously insisted on the indispensability of what he termed ``classical concepts.'' In the context of the decoherence program, on the other hand, it has become fashionable to talk about the ``dynamical emergence of classicality'' from the quantum formalism alone. Does this mean that decoherence challenges Bohr's dictum---for example, that classical concepts do not need to be assumed but can be derived? In this paper, we'll try to shed some light down the murky waters where formalism and philosophy mingle. To begin, we'll clarify the notion of classicality in the decoherence description. We'll then discuss Bohr's and Heisenberg's takes on the quantum--classical problem and reflect on the different meanings of the terms ``classicality'' and ``classical concepts'' in the writings of Bohr and his followers. This analysis will allow us to put forward some tentative suggestions for how we may better understand the relation between decoherence-induced classicality and Bohr's classical concepts.
\end{abstract}

\maketitle

\section{Introduction}

What does the term ``classical'' mean in the context of quantum mechanics? A wealth of confusing uses of this word abounds in the literature. For instance, the formalism of quantum mechanics is often said to arise from a quantization of classical variables. And while the textbook axioms make no mention of the word ``classical,'' there's often talk of a classical limit, as in the context of Bohr's correspondence principle, Ehrenfest's theorem, or Feynman's path-integral formalism.

And turning to more philosophically tinged territory, Niels Bohr famously insisted on the indispensability of classical concepts in applying and making sense of quantum theory. In particular, he assigned paramount importance to the notion of classical measuring apparatuses---a move that served as a lightning rod for countless foundational debates and may have fueled a fair amount of anti-Copenhagen rhetoric.

Now enter the modern age of quantum theory, and with it one of its darlings: the decoherence program \cite{Joos:2003:jh,Zurek:2002:ii,Schlosshauer:2003:tv,Schlosshauer:2007:un}. Decoherence has boldly proclaimed its ability to explain the emergence of ``classicality'' and the ``classical world of our experience'' from within quantum mechanics. And decoherence declares to accomplish this goal without recourse to any additional formal or philosophical baggage. 

``What else could we need!'' we exclaim with delight. ``Clearly, decoherence makes the \emph{a priori} assumption of classical structures superfluous. Bohr was wrong!''

After all, so our thinking might go, if classicality can be shown to arise as a natural consequence of a realistic application of the quantum formalism, how could Bohr's insistence on the irreducibility of the classical concepts possibly remain justified and stand any chance at all? 

Indeed, many people seem to feel that way. For example, Zeh has contrasted the dynamical approach of decoherence to the quantum--classical problem with what he calls the ``irrationalism'' of the Copenhagen school \cite[p.~27]{Joos:2003:jh}. There's always been a good deal of resentment to Bohr's mantra of classical concepts, and decoherence appears to have only aggravated these feelings.

While it might be tempting to read the implications of the decoherence mechanism as making Bohr's dictum redundant, in this paper we shall argue that the matter is in fact significantly more subtle. To anticipate our conclusion in one oversimplified, informal slogan: \emph{Bohr and decoherence aren't talking about the same kind of classicality}.

Only once such differences are properly recognized can we hope to understand how the decoherence account and Bohr's classical concepts relate to each other and may challenge each other. In this paper, we'll make an attempt in this direction. Our analysis won't be the last word, but it's a start. We shall also refer you to a previous (and significantly longer) paper of ours \cite{Schlosshauer:2008:im}. There, liberated from the space constraints of a conference paper, we devote much more room to the fascinating historical aspects of the subject. We're just mentioning this tidbit in case you're experiencing an insatiable hunger for a second helping of quotes from Bohr and his disciples. If not, then this paper may be enough to quench your thirst.

\section{Classicality in the\\ decoherence program}

Decoherence is not an addition to quantum theory but simply a consequence of it. Its formal description proceeds within the standard framework of quantum states and density matrices. Let's begin by recapping the decoherence formalism in all brevity (see \cite{Joos:2003:jh,Zurek:2002:ii,Schlosshauer:2003:tv,Schlosshauer:2007:un} for actual reviews).

We're considering two parts, a system and its environment, and a von Neumann measurement-like interaction between the two entities. This makes the superposition, which had been initially confined to the system, spread to the environment, and we end up with an entangled composite state. We're now interested in what we may be able to observe by measuring the system, disregarding the environment. This restriction simply reflects the fact that, in practice, we can't actually access the majority of the environmental degrees of freedom. Quantum mechanics tells us how to compute the statistics of all possible measurements on the system: we need to simply ``trace out'' the environment in the system--environment density matrix. And what pops out is the reduced density matrix, which is the main object of interest for decoherence. 

What we now find is that typically this density matrix will be approximately diagonal in an eigenbasis of some classical observable, such as position, and that the density matrix will remain approximately diagonal over time. (The specific basis is determined by the structure of the interaction Hamiltonian.) This means that interference terms in this basis become quickly and strongly damped. Such terms correspond to (the probabilities of) possible outcomes of measurements of observables that are often explicitly nonclassical. It is in this sense that decoherence explains why it is usually easy to measure a classical observable but so painfully difficult to observe nonclassical states.

So let's note two things at this point. First, the classicality that decoherence is talking about is strictly associated with properties of reduced density matrices. Decoherence shows how the evolution of the system's (reduced) density matrix is influenced by the system's coupling to its environment. Decoherence thus amounts to a filter on the space of density matrices, singling out those density matrices that can be practically realized given the inevitably openness of any realistic physical system. 

And this immediately leads us to our second point. When we're focusing our search for classicality to level of the quantum-state formalism---as decoherence is doing---then the kind of states decoherence deems classical in appearance are in fact, arguably, among the least classical ones we can think of. To see why, it'll be helpful to sketch a hierarchy of classicality for quantum states. 

A natural definition of a classical observable would be an observable that corresponds to the measurement of a quantity that exists in classical physics. Energy, angular momentum, position, and momentum are obvious examples. However, the degree of classicality will depend on whether the observable refers to discrete or continuous quantities. A pure eigenstate of a classical observables with a discrete spectrum would be the most classical quantum state: the system then possesses a well-defined, sharp value of a classical physical quantity. 

Many classical observables, on the other hand, have a continuous spectrum. Position and momentum are textbook examples. It is well known that there are no physically meaningful quantum states that are sharp eigenstates of continuous observables, since such states would be non-normalizable. Therefore the best we can do is to form narrow wave packets (e.g., Gaussians) in these variables, that is, superpositions of (say) eigenkets $\ket{x}$ or $\ket{p}$ with a narrow spread in $x$ or $p$, or in both directions (coherent states are those optimally narrow in both $x$ and $p$). Such states are often called ``quasiclassical.'' This term is quite appropriate, though it may not fully express the strong quantum--classical tension inherent in a state of this kind. On the one hand, a (say) coherent state appears indeed similar to that of a classical point mass with a somewhat smeared-out phase-space trajectory. On the other hand, however, the state remains distinctly quantum-mechanical, because it is a coherent superposition of the ``classical'' state vectors $\ket{x}$ (or $\ket{p}$). So approximate eigenstates of continuous classical observables will obviously rank lower on the classicality scale than sharp eigenstates of discrete classical observables. 

What about observables without counterpart in classical physics? A mild intermediate case is spin. Spin is a truly quantum-mechanical quantity, and therefore spin observables are not classical per se. But one may say that spin is largely analogous to the concept of angular momentum in classical physics, and it is in this sense that spin observables are often regarded as residing on a similar footing as classical observables. 

But this peculiar example aside, pretty much any observable that the quantum formalism allows us to write down fails to represent anything familiar from classical physics. A particular class are ``proverbially nonclassical'' observables, that is, observables whose eigenstates correspond to a superposition of macroscopically distinct eigenstates of classical observables. One (if somewhat silly) example is the projective observable directly verifying the presence of a Schr\"odinger-cat superposition. It should be obvious that eigenstates of such observables are located at the very bottom of the classicality scale of quantum states. We note in passing that it is precisely such nonclassical, counterintuitive cat-type states that are often used to illustrate the consequences of decoherence. 

So far, we've only talked about pure states. A proper mixture $\op{\rho} = \sum_k p_k \ketbra{\psi_k}{\psi_k}$ won't introduce anything spectacularly new, because such a mixture simply represents an ignorance-interpretable \cite{Espagnat:1988:cf} ensemble, i.e., the coefficients $p_k$ represent probabilities that express purely our \emph{subjective} ignorance about the particular pure state $\ket{\psi_k}$ the system has actually been prepared in.

However, the creatures decoherence toys with are of a very different breed. Because the system is caught up in massive environmental entanglement, we can no longer work with pure states or proper mixtures: reduced density matrices is all we've got at our disposal. These are improper mixtures, where the term ``improper'' is used to signify that, all apparent formal similarities notwithstanding, such mixtures must not be interpreted as an ignorance-interpretable (proper) ensemble. Instead, reduced density matrices do no more---and no less---than encapsulate the statistics of all future local measurements that could be performed on the system. Because of entanglement, no quantum state can be assigned to the system alone, and improper density matrices cannot be used to infer a classical ensemble of states of which one state is actually realized in the system. 

It should now be clear why the states that arise from decoherence rank so low on the classicality scale of quantum states. These states are reduced density matrices describing improper ensembles. They evolve toward only approximate diagonality, leaving nonvanishing interference terms. And the eigenbasis is often merely quasiclassical, as in the case of improper mixtures of coherent states. This is a far cry from what we'd consider an obviously classical quantum state, such as a pure sharp eigenstate of a classical observable. But needless to say, this statement is meant as a clarification, not as a critical jab at decoherence: the kind of states decoherence gives us are \emph{as good as they get}, given the inevitable openness of realistic quantum systems.

\section{Classicality in Bohr's and Heisenberg's interpretation of quantum mechanics}

It is quite fashionable to contrast Bohr's we-need-classical-measuring-apparatuses account with von Neumann's just-apply-the-quantum-formalism-to-everything description of measurement. In this reading, it is as if Bohr had been desperately clinging to some vestige of a classical mindset, whereas von Neumann had been the admirably consistent all-is-quantum man who let the quantum do the talking. 

But Bohr recognized very well the dramatic consequences of nonseparability resulting from entanglement---as starkly illustrated by von Neumann's measurement scheme---and viewed this nonseparability as a central epistemological paradox posed by quantum theory. Bohr explained that, in contrast with classical physics, quantum mechanics confronts us with the ``impossibility of any sharp separation between the behavior of atomic objects and the interaction with the measuring instruments which serve to define the very conditions under which the phenomena appear'' \cite[p.~210]{Bohr:1949:mz}. If the system and the measuring probe become an entangled quantum-mechanical whole, then the distinction between system and probe becomes fundamentally ambiguous. For Bohr, the possibility of such a distinction was a logical necessity for empirical inquiry. If everything is just gobbled up by ever-spreading entanglement and homogenized into one gargantuan maelstrom of nonlocal quantum holism, and if we can't conceptually isolate and localize a system and regard it as causally independent from some (potentially distant) other system, then there are no systems that could be the object of empirical knowledge. In order to meaningfully speak of observation in quantum mechanics, Bohr concluded, ``one must therefore cut out a partial system somewhere from the world, and one must make `statements' or `observations' just about this partial system'' \cite[p.~141]{Bohr:1985:mn}. And so, while quantum mechanics \emph{could} certainly be used to describe the interaction between system and the measuring instrument, as von Neumann explicitly did, for Bohr such a move was self-defeating, because to do so would render the distinction between object and instrument ambiguous and preclude one from treating the measuring instrument as such. Heisenberg, in discussions that followed Bohr's Como paper in 1927, expressed a similar sentiment: ``One may treat the whole world as \emph{one} mechanical system, but then only a mathematical problem remains while access to observation is closed off'' \cite[p.~141]{Bohr:1985:mn}. 

Bohr's epistemological demand for a ``cut between the observed system on the one hand and the observer and his apparatus on the other hand'' also became a key theme of Heisenberg's thinking. However, for Heisenberg, \emph{the object--instrument divide was coincident with the quantum--classical divide}. For instance, in a 1934 lecture, Heisenberg emphasized that ``there arises the necessity to draw a clear dividing line in the description of atomic processes, between the measuring apparatus of the observer which is described in classical concepts, and the object under observation, whose behavior is represented by a wave function'' \cite[p.~15]{Heisenberg:1952:mn}. The following year, he elaborated on this idea:
\begin{quote}
  In this situation it follows automatically that, in a mathematical
  treatment of the process, a dividing line must be drawn between, on
  the one hand, the apparatus which we use as an aid in putting the
  question and thus, in a way, treat as part of ourselves, and on the
  other hand, the physical systems we wish to investigate. The latter
  we represent mathematically as a wave function. This function,
  according to quantum theory, consists of a differential equation
  which determines any future state from the present state of the
  function \dots\ The dividing line between the system to be observed
  and the measuring apparatus is immediately defined by the nature of
  the problem but it obviously \emph{signifies no discontinuity of the
    physical process}. For this reason there must, within certain
  limits, exist complete freedom in choosing the position of the
  dividing line \cite[p.~49, emphasis added]{Heisenberg:1952:mk}.
\end{quote}
First, it's interesting to note that Heisenberg conjures up the image of the apparatus as something that's part of the observer, as a kind of prosthetic hand if you wish (see the program of Quantum Bayesianism \cite{Fuchs:2010:az} for a modern take on this idea). Second, the italicized phrase of the quote clearly shows that the cut ``obviously'' must not be understood as a physical boundary: it's not a border at which the laws of physics change, or something that delineates the breakdown region of quantum mechanics. Third, as the above quote hints at, for Heisenberg the ``cut can be shifted arbitrarily far in the direction of the observer in the region that can otherwise be described according to the laws of classical physics,'' but, of course, ``the cut cannot be shifted arbitrarily in the direction of the atomic system'' \cite[p.~414]{Heisenberg:1985:zq}. The cut ``cannot be established physically''---it represents no physical discontinuity---``and moreover it is precisely the arbitrariness in the choice of the location of the cut that is decisive for the application of quantum mechanics'' \cite[p.~416]{Heisenberg:1985:zq}. 

It is important to realize that Heisenberg's views on the cut were somewhat at odds with Bohr's own view of the problem. Indeed, in an exchange of correspondence in 1935 Heisenberg and Bohr argued the point without resolution. Some twenty years later Heisenberg would report that ``Bohr has emphasized that it is more realistic to state that the division into the object and rest of the world is not arbitrary'' and that the object is determined by the very nature of the experiment \cite[p.~24]{Heisenberg:1989:zb}. In a letter to Heelan in 1975, Heisenberg also explained that he and Bohr had never really resolved their disagreement. Heisenberg remained convinced ``that a cut could be moved around to some extent while Bohr preferred to think that the position is uniquely defined in every experiment'' \cite[p.~137]{Heelan:1975:kk}. While a closer analysis of the nature of this disagreement would take us well beyond the scope of this paper, this episode serves to remind us that we should be careful about attributing the views of Bohr's contemporaries to Bohr himself. 

But let's move on and take a closer look at Bohr's classical concepts. What exactly is their nature? And why are we entitled to use them at all, given that the laws of quantum mechanics trump those of classical physics? To shed some light on these questions, and to clarify possible connections with decoherence, it's useful to make two points. The first is that for Bohr, quantum mechanics makes use of classical concepts, such as position and momentum, \emph{in spite of the limitations of their applicability}. Thus, what we earlier called ``quasiclassical'' states were, for Bohr, evidence of the continued use of concepts borrowed from classical physics. As Heisenberg recalled from his discussions with Bohr in 1927: ``Well, in spite of your Uncertainty Principle you have got to use words like `position' and `velocity' just because you haven't got anything else'' \cite[interview with Thomas S.\ Kuhn, 27 February 1963]{AHQP:1986:po}. This is why Bohr repeatedly insisted that ``it lies in the nature of physical observation \dots\ that all experience \emph{must} ultimately be expressed \emph{in terms of} classical concepts'' \cite[p.~94, emphasis added]{Bohr:1987:aw}. 

But why do we, or why must we, interpret observations in classical terms? One approach amounts to an epistemological formulation of the doctrine of classical concepts. Its aim was to elucidate why our conceptual framework is so wedded to our classical intuitions about the world. Such accounts are often given a decisively linguistic spin. For example, Petersen suggested that Bohr's remarks on the indispensability of classical concepts ``are based on his general attitude to the epistemological status of language and to the meaning of unambiguous conceptual communication, and they should be interpreted in that background'' \cite[p.~179]{Petersen:1968:uu}. The following quote from Bohr---a rather clear expression of his doctrine---would appear to support such a reading:
\begin{quote}
  It is decisive to recognize that, \emph{however far the phenomena
    transcend the scope of classical physical explanation, the account
    of all evidence must be expressed in classical terms.}  The
  argument is simply that by the word ``experiment'' we refer to a
  situation where we can tell others what we have done and what we
  have learned and that, therefore, the account of the experimental
  arrangement and of the results of the observations must be expressed
  in unambiguous language with suitable application of the terminology
  of classical physics \cite[p.~209]{Bohr:1949:mz}.
\end{quote}

This brings us to the second point concerning Bohr's views. Passages such as this one have often been interpreted as reflecting Bohr's attitude that we are ``suspended in language.'' At the same time, however, we must pay particular attention to the emphasis that Bohr placed on the purpose of experiment. In Bohr's view, any attempt to account for the \emph{epistemic function} of an experiment and for the way the experiment has been \emph{designed} will invariably fall back on the use of classical concepts. While the measuring instrument may of course be described as a quantum-mechanical system, it is only possible to explain its capacity to function \emph{as} an instrument on the assumption that the interaction between object and instrument can be described in terms of an exchange of energy and momentum somewhere in space and time. This is why, for Bohr, ``the unambiguous interpretation of any \emph{measurement} must be essentially framed in terms of classical physical theories, and we may say that in this sense the language of Newton and Maxwell will remain the language of physics for all time'' \cite[p.~692, emphasis added]{Bohr:1931:ii}. 

It is important to bear in mind that Bohr's doctrine of classical concepts was reinterpreted by many of his contemporaries. For instance, while Bohr expressed the view that we \emph{must} use classical concepts, Bohr disciples, such as Rosenfeld, Heisenberg, and Weizs\"acker, tended to prefer a less dogmatic reading. In their hands, the doctrine was transformed from a categorical imperative to a pragmatic statement of fact (see also \cite[pp.~197--9]{Beller:1999:za}). Witness, for example, this quote from Weizs\"acker:
\begin{quote}
  We ought not to say, ``Every experiment that is even possible
  \emph{must} be classically described,'' but ``Every actual
  experiment known to us \emph{is} classically described, and we do
  not know how to proceed otherwise.'' \dots\ 
  Thus the factual, we might almost say historical situation of
  physics is made basic to our propositions \cite[pp.~128,
  130]{Weizsacker:1952:ym}.
\end{quote}
This later, more pragmatic, position would seem to sit better with many physicists today. At the end of the day, however, we're still left with the question of why is the world is wired in such a way that the concepts of classical physics are applicable at all. This entails that we shift from an \emph{epistemological} formulation of the question (characteristic of Bohr's own approach) to an \emph{ontological} one. This second approach was indeed pursued in the 1950s and 1960s by some of Bohr's acolytes. They were seeking a physical formulation of the doctrine of classical concepts that would account for the emergence of some kind of effective classicality from the quantum description. Weizs\"acker described the task at hand:
\begin{quote}
Having thus accepted the falsity of classical physics, taken literally, we must ask how it can be explained as an essentially good approximation [when describing objects at the macrolevel]. \dots\ This amounts to asking \emph{what physical condition must be imposed on a quantum-theoretical system in order that it should show the features which we describe as ``classical.''} \dots\ I am unable to prove mathematically that the condition of irreversibility would suffice to define a classical approximation, but I feel confident it is a necessary condition \cite[pp.~28--29, emphasis in original]{Weizsacker:1971:ll}.
\end{quote}
The idea that an irreversibility of the processes taking place in the apparatus should play a key role was also picked up by Rosenfeld, who subsequently advocated---in an ``unqualified endorsement,'' in Jammer's words \cite[p.~493]{Jammer:1974:pq}---the proposal of Daneri, Loinger, and Prosperi \cite{Daneri:1962:om}. At any rate, curiously both Weizs\"acker and Rosenfeld defended their physical approach to the quantum-to-classical transition as being in line with the spirit in which Bohr had intended his doctrine of classical concepts.

\section{The relationship between decoherence and the doctrine of classical concepts} 

To what extent does decoherence mark a departure from the doctrine of classical concepts? As our analysis should have made clear, the answer will depend on the particular interpretation of the doctrine. Decoherence and Bohr's own reading of classical concepts plow rather different fields. Decoherence is concerned with the evolution of reduced density matrices and describes the dynamical selection of (improper) ensembles of certain, typically ``quasiclassical'' quantum states. Bohr's doctrine of classical concepts, on the other hand, is never phrased in terms of quantum states; it is about interpreting quantum theory. It is motivated by two demands that Bohr saw as a prerequisite for an unambiguous and objective description of quantum phenomena. First, the logical demand for separability between the observed (the system) and the observer (the apparatus). And second, the demand for a classical description of the functioning of the measuring instrument. Bohr's doctrine is couched in categorical terms, and given an epistemological or even linguistic reading. 

Since decoherence is simply a consequence of a realistic application of the standard quantum formalism, it cannot by itself give an interpretation or explanation of this formalism. Bohr's fundamental point was that any interpretation of quantum mechanics must in the end fall back on the use of classical concepts. And this means that, while we may be inclined to invoke decoherence to justify, in a practical sense, Bohr's philosophical stance about the categorical use of classical concepts, we would run a certain risk of circularity in making such an argument.

It is the rather different versions of the doctrine advocated and developed by Bohr's followers---in particular, the physical and pragmatic approaches of Rosenfeld, Weizs\"acker, and others---that make closer contact with the spirit of the decoherence program and its notion of classicality. Decoherence may here be seen as providing a physical justification for the pragmatic use of classical concepts in a given experimental situation. And \emph{if} we choose to \emph{re}-interpret the Heisenberg cut as the boundary at which it becomes difficult to observe certain quantum states, then decoherence explains and quantifies the cut. In this interpretation, then, decoherence explains why, when, and where classical physics is a good approximation in dealing with the quantum world, thus providing an answer to the question that Heisenberg, Rosenfeld, and Weizs\"acker were asking when they tried to find a physical foundation for Bohr's doctrine.

Howard \cite{Howard:1994:lm}, in his ``reconstruction'' of Bohr's interpretation, has suggested that Bohr's classical concepts may be identified with the selection of subensembles that are appropriate to the measurement context (or, in Bohr's terminology, that are appropriate to the particular ``experimental arrangement''). That is, to apply classical concepts means replacing the global pure entangled system--apparatus state (resulting from von Neumann's measurement scheme) with the proper mixture for the measured system expressed in the eigenbasis of the to-be-measured observable. This mixture then exhaustively encapsulates the statistics for the chosen measurement. Decoherence could be nicely put to use in such account, as it would describe the dynamical emergence of such (albeit improper!) mixtures. 

Interestingly, one can find a few passages in Heisenberg's writings that seem to hint at an early recognition of the role of the environment---which is so paramount to the decoherence program---in the relationship between quantum and classical. Here's an example, from \emph{Physics and Philosophy}:
\begin{quote}
  It must be observed that the system which is treated
  by the methods of quantum mechanics is in fact a part of a much
  bigger system (eventually the whole world); it is interacting with
  this bigger system; and one must add that the microscopic properties
  of the bigger system are (at least to a large extent) unknown. This
  statement is undoubtedly a correct description of the actual
  situation \dots\ The interaction with the bigger system with its
  undefined microscopic properties then introduces a new statistical
  element into the description \dots\ of the system under
  consideration. In the limiting case of the large dimensions this
  statistical element destroys the effects of the ``interference of
  probabilities'' in such a manner that the quantum-mechanical scheme
  really approaches the classical one in the limit
  \cite[pp.~121--2]{Heisenberg:1989:zb}.
\end{quote}
And elsewhere, Heisenberg argues that the quantum-to-classical transition depends on ``the underlying assumption,'' implicit in the Copenhagen interpretation, ``that the interference terms are in the actual experiment removed by the partly undefined interactions of the measuring apparatus, with the system and with \emph{the rest of the world} (in the formalism, the interaction produces a `mixture')'' \cite[p.~23]{Heisenberg:1955:lm}. This comes surprisingly close to the spirit of decoherence. However, one should not overinterpret Heisenberg's clairvoyance: the crucial ingredient of decoherence, entanglement, is not part of Heisenberg's account.

What would Bohr say if he had known about decoherence? Our feeling is that, while he would have certainly regarded decoherence as a useful application of quantum theory, he would probably not have changed his position on the irreducibility of classical concepts. Indeed, as we hope to have shown, maintaining a peaceful coexistence between Bohr's philosophy and decoherence may be more feasible than often claimed. 

\begin{acknowledgments}
M.S.\ acknowledges financial support from the Danish Research Council.
\end{acknowledgments}

\bibliographystyle{aipproc}   

\begin{thebibliography}{26}
\expandafter\ifx\csname natexlab\endcsname\relax\def\natexlab#1{#1}\fi
\providecommand{\enquote}[1]{``#1''}
\expandafter\ifx\csname url\endcsname\relax
  \def\url#1{\texttt{#1}}\fi
\expandafter\ifx\csname urlprefix\endcsname\relax\def\urlprefix{URL }\fi
\providecommand{\eprint}[2][]{\url{#2}}

\bibitem[Joos et~al.(2003)]{Joos:2003:jh}
E.~Joos, H.~D. Zeh, C.~Kiefer, D.~Giulini, J.~Kupsch, and I.-O. Stamatescu,
  \emph{Decoherence and the Appearance of a Classical World in Quantum Theory},
  Springer, New York, 2003, 2nd edn.

\bibitem[Zurek(2003)]{Zurek:2002:ii}
W.~H. Zurek, \emph{Rev. Mod. Phys.} \textbf{75}, 715--775 (2003).

\bibitem[Schlosshauer(2004)]{Schlosshauer:2003:tv}
M.~Schlosshauer, \emph{Rev. Mod. Phys.} \textbf{76}, 1267--1305 (2004).

\bibitem[Schlosshauer(2007)]{Schlosshauer:2007:un}
M.~Schlosshauer, \emph{Decoherence and the Quantum-to-Classical Transition},
  Springer, Berlin/Heidelberg, 2007, 1st edn.

\bibitem[Schlosshauer and Camilleri(2008)]{Schlosshauer:2008:im}
M.~Schlosshauer, and K.~Camilleri  (2008), \eprint{arXiv:0804.1609 [quant-ph]}.

\bibitem[{d'E}spagnat(1976)]{Espagnat:1988:cf}
B.~{d'E}spagnat, \emph{Conceptual Foundations of Quantum Mechanics}, Benjamin,
  Reading, Massachusetts, 1976, 2nd edn.

\bibitem[Bohr(1949)]{Bohr:1949:mz}
N.~Bohr, \enquote{Discussions with {E}instein on Epistemological Problems in
  Atomic Physics,} in \emph{Albert Einstein: Philosopher--Scientist}, edited by
  P.~A. Schilpp, Library of Living Philosophers, Evanston, Illinois, 1949, pp.
  201--241, reprinted in {\cite{Wheeler:1983:bc}}, pp.~9--49.

\bibitem[Bohr(1985)]{Bohr:1985:mn}
N.~Bohr, \enquote{Collected Works,} in \emph{Vol.~6. Foundations of Quantum
  Mechanics~I (1926--1932)}, edited by J.~Kalckar, North-Holland, Amsterdam,
  1985.

\bibitem[Heisenberg(1952{\natexlab{a}})]{Heisenberg:1952:mn}
W.~Heisenberg, \enquote{Recent changes in the foundations of exact science,} in
  \emph{Philosophic Problems in Nuclear Science}, Faber and Faber, London,
  1952{\natexlab{a}}, pp. 11--26, translated by F. C. Hayes.

\bibitem[Heisenberg(1952{\natexlab{b}})]{Heisenberg:1952:mk}
W.~Heisenberg, \enquote{Questions of principle in modern physics,} in
  \emph{Philosophic Problems in Nuclear Science}, Faber and Faber, London,
  1952{\natexlab{b}}, pp. 41--52, translated by F. C. Hayes.

\bibitem[Fuchs(2010)]{Fuchs:2010:az}
C.~A. Fuchs  (2010), \eprint{arXiv:1003.5209v1 [quant-ph]}.

\bibitem[Heisenberg(1985)]{Heisenberg:1985:zq}
W.~Heisenberg, \enquote{Ist eine deterministische {E}rg{\"a}nzung der
  {Q}uantenmechanik m{\"o}glich?,} in \emph{Wolfgang Pauli. Wissenschaftlicher
  Briefwechsel mit Bohr, Einstein, Heisenberg}, edited by A.~Hermann, K.~von
  Meyenn, and V.~F. Weisskopf, Springer, New York, 1985, vol. 1: 1919--1929,
  pp. 409--418.

\bibitem[Heisenberg(1989)]{Heisenberg:1989:zb}
W.~Heisenberg, \emph{Physics and Philosophy. The Revolution in Modern Science},
  Penguin, London, 1989.

\bibitem[Heelan(1975)]{Heelan:1975:kk}
P.~Heelan, \emph{Z. Allgemeine Wissenschaftstheorie} \textbf{6}, 113--138
  (1975).

\bibitem[AHQ(1986)]{AHQP:1986:po}
\emph{Archives for the History of Quantum Physics}, American Philosophical
  Society, Philadelphia, 1986, 301 microfilm reels.

\bibitem[Bohr(1987)]{Bohr:1987:aw}
N.~Bohr, \emph{Atomic Theory and the Description of Nature. The Philosophical
  Writings of Niels Bohr}, vol.~1, Ox Box, Woodbridge, Connecticut, 1987.

\bibitem[Petersen(1968)]{Petersen:1968:uu}
A.~Petersen, \emph{Quantum Physics and the Philosophical Tradition}, MIT Press,
  Cambridge, 1968.

\bibitem[Bohr(1931)]{Bohr:1931:ii}
N.~Bohr, \emph{Nature} \textbf{128}, 691--692 (1931).

\bibitem[Beller(1999)]{Beller:1999:za}
M.~Beller, \emph{Quantum Dialogue: The Making of a Revolution}, University of
  Chicago, Chicago, 1999.

\bibitem[Weizs{\"a}cker(1952)]{Weizsacker:1952:ym}
C.~F.~v. Weizs{\"a}cker, \emph{The World View of Physics}, Routledge, London,
  1952, translated by M. Grene.

\bibitem[Weizs{\"a}cker(1971)]{Weizsacker:1971:ll}
C.~F.~v. Weizs{\"a}cker, \enquote{The {C}openhagen interpretation,} in
  \emph{Quantum Theory and Beyond}, edited by T.~Bastin, Cambridge University,
  Cambridge, 1971, pp. 25--31.

\bibitem[Jammer(1974)]{Jammer:1974:pq}
M.~Jammer, \emph{The Philosophy of Quantum Mechanics}, John Wiley \& Sons, New
  York, 1974, 1st edn.

\bibitem[Daneri et~al.(1962)]{Daneri:1962:om}
A.~Daneri, A.~Loinger, and G.~M. Prosperi, \emph{Nucl. Phys.} \textbf{33},
  297--319 (1962).

\bibitem[Howard(1994)]{Howard:1994:lm}
D.~Howard, \enquote{What Makes a Classical Concept Classical? {T}oward a
  Reconstruction of {N}iels {B}ohr's Philosophy of Physics,} in \emph{Niels
  Bohr and Contemporary Philosophy}, Kluwer, Dordrecht, 1994, vol. 158 of
  \emph{Boston Studies in the Philosophy of Science}, pp. 201--229.

\bibitem[Heisenberg(1955)]{Heisenberg:1955:lm}
W.~Heisenberg, \enquote{The Development of the Interpretation of the Quantum
  Theory,} in \emph{Niels Bohr and the Development of Physics: Essays Dedicated
  to Niels Bohr on the Occasion of his Seventieth Birthday}, edited by
  W.~Pauli, L.~Rosenfeld, and V.~Weisskopf, McGraw Hill, New York, 1955, pp.
  12--29.

\bibitem[Wheeler and Zurek(1983)]{Wheeler:1983:bc}
J.~A. Wheeler, and W.~H. Zurek, editors, \emph{Quantum Theory and Measurement},
  Princeton University, Princeton, 1983.

\end{thebibliography}

\end{document}